# A Simple Relativity Solution to the Bell Spaceship Paradox

Ralph Berger, PhD, University of California at Berkeley

Abstract

The Bell Spaceship Paradox has promoted confusion and numerous resolutions since its first statement in 1959, including resolutions based on relativistic stress due to Lorentz contractions. The paradox is that two ships, starting from the same reference frame and subject to the same acceleration, would snap a string that connected them, even as their separation distance would not change as measured from the original reference frame. This paper uses a Simple Relativity approach to resolve the paradox and explain both why the string snaps, and how to adjust accelerations to avoid snapping the string. In doing so, an interesting parallel understanding of the Lorentz contraction is generated. The solution is applied to rotation to address the Ehrenfest paradox and orbital precession as well.

Introduction

The Bell Spaceship Paradox dates from 1959 [1], and describes a thought experiment involving two spaceships, initially at rest in an inertial frame, one in front of the other, and tied with a string. The spaceships then both accelerate at a constant rate, say 1g. The paradox is that, though both ships have the same acceleration at any moment as calculated from the original reference frame, the string between them will be stretched and broken. One visualization is that the length of the string must contract by the familiar Lorentz relation [25]

$$L/L_o = 1/\gamma = \sqrt{1 - (v/c)^2}, \qquad [1]$$

yet the distance between the ships is constant as measured from the original reference frame.

A large number of papers have been published on this paradox, and the current Wikipedia entry lists [2] Evett & Wangsness (1960), [3] Dewan (1963),[4] Romain (1963),[5] Evett (1972),[6] Gershtein & Logunov (1998),[7] Tartaglia & Ruggiero (2003),[8] Cornwell (2005),[9] Flores (2005),[10] Semay (2006),[11] Styer (2007),[12] Freund (2008),[13] Redzic (2008),[14] Peregoudov (2009),[15] Redžić (2009),[16] Gu (2009),[17] Petkov (2009),[18] Franklin (2009),[19] Miller (2010),[20] Fernflores (2011),[21] Kassner (2012),[22] Natario (2014),[23] Lewis, Barnes & Sticka (2018), and [24] Bokor (2018). These papers include explanations, mostly consistent with Bell's own, that the string breaks due to relativistic stress. Bell [1] described it as a physical effect that can be explained with relativistic electromagnetism. However, two disagree (Petrov [18] and Franklin [19]) noting Lorentz length contraction has no "physical reality" and the string breaks only because of "a rotation in four dimensional space which by itself can never cause any stress at all."

Adding to the confusion, should a single spaceship of length equal to the string be accelerated, it would not be subject to any length-change stress. Passengers on such a spacecraft would never feel that its length is contracted, even as we still observers would think Equation 1 applied.

Simple relativity is an approach that attempts to explain complex relativity phenomena in a simple manner. Many of the odd effects of relativity, such as planetary precession or light path deflection, can be explained in very simple terms using conservation of energy and modifications of Newtonian laws. The Bell Spaceship paradox can be solved in this way, which also provides the spaceship captains with an acceleration scheme which they could adopt to avoid breaking the string. If fact, simple relativity makes this difficult paradox into a simple and obvious thing.

Equivalence principle

One criticism of the string-breaks theory is that the 1g acceleration due to thrust ought to be equivalent to a 1g acceleration due to planetary gravity. If a string is tied from one point to another 1000 m above it, the string develops no stress even if left in that acceleration for a year. Why should a string connecting two spaceships each accelerating at 1g develop any internal stress? Is that a violation of the equivalence principle?

But no, we see a problem with that analogy. The acceleration of a point 1000m above another on Earth is not identical. Acceleration due to gravity is

$$g = GM/r^2 \qquad [2]$$

where $G$ is the universal gravitational constant, $M$ the planet mass, and $r$ the distance to the planet's center. Hence it is not true that points that differ by vertical distances on Earth are subject to the same acceleration; a higher point is at a lower acceleration. If we tried to impose the same acceleration, then clearly the higher point would rise, and the string break.

Time discrepancies within a spaceship

We note that time speed differences occur in planetary gravitation fields with elevation. Einstein's rock-falling-in-a-well thought experiment explains why, using conservation of energy. The idea is that a rock falls, gaining velocity v and apparent mass. At the bottom of the well, an assistant uses $E=mc^2$ to convert the rock to a photon with equivalent energy $hf$ (Planck's constant times frequency) and shines it back to the surface. There our rock-dropper converts it back to mass. To avoid any conflicts with conservation of mass and energy, the final mass must be identical to the initial. That only happens if the frequency was reduced during the upward flight. Using the escape velocity as fall velocity (to allow a relative time speed to a distant viewer outside of the gravitational potential), we replace the $v^2$ term of Equation 1 with the square of escape velocity 2GM/r to get a time ratio due to gravitational potential of:

$$t/t_o = 1/\gamma_g = \sqrt{1 - 2GM/rc^2} = \sqrt{1 - r_s/r} \qquad [3]$$

The difference in time speed for a 1000 m height on Earth can be found using Equation 3 recognizing that, for Earth, $2GM/c^2$ also written above as $r_s$ is a value of 0.0088681 meters:

$$\Delta\gamma_g = 1\Big/\sqrt{1 - 0.0088681/6.374e6} - 1\Big/\sqrt{1 - 0.0088681/6.375e6} = 1.0925e - 13$$

A similar time discrepancy exists inside an accelerating spaceship. As we can use the rock-falling-into-a-well though experiment to develop Equation 3, we can use conservation of energy to develop a similar result for a tall spaceship of height $h$ using kinetic energy $1/2mv^2 = mgh$ for all practical fall velocities and rocket ship accelerations. Putting the fall speed $v^2 = 2gh$ into the time dilation formula Equation 1 gives us a change in time speed for fall distance of $h$:

$$\Delta\gamma = 1\bigg/\sqrt{1 - \frac{2gh}{c^2}} - 1 \qquad [4]$$

where we are using 1 as the time speed at the top of the spaceship where $h$ is zero.

For our 1g, 1000 m example, Equation 4 gives a time ratio that is identical to the Earthly example:

$$\Delta\gamma = 1\bigg/\sqrt{1 - \frac{2*9.81*1000}{c^2}} - 1 = 1.0925e - 13$$

Hence time runs a little faster for the tip of a spaceship than it does for the tail, just as time runs a little faster on the summit of a mountain as compared to the base.

Aside: time speed changes on planets depend on the density, but unless the planet is extremely dense, the effects of time speed versus elevation as a function of distance is pretty much the same for a planet as for a spaceship. Plotting Equations 3 and 4 for planets with surface acceleration of 9.81 m/s² gives the following result. It may seem counter-intuitive that the high density planet has less change in time speed per altitude, but the reason is that the time speed change is related to fall velocity, and the smaller radius, dense planets have a more rapid fall off in gravitational acceleration with altitude, hence less fall velocity.

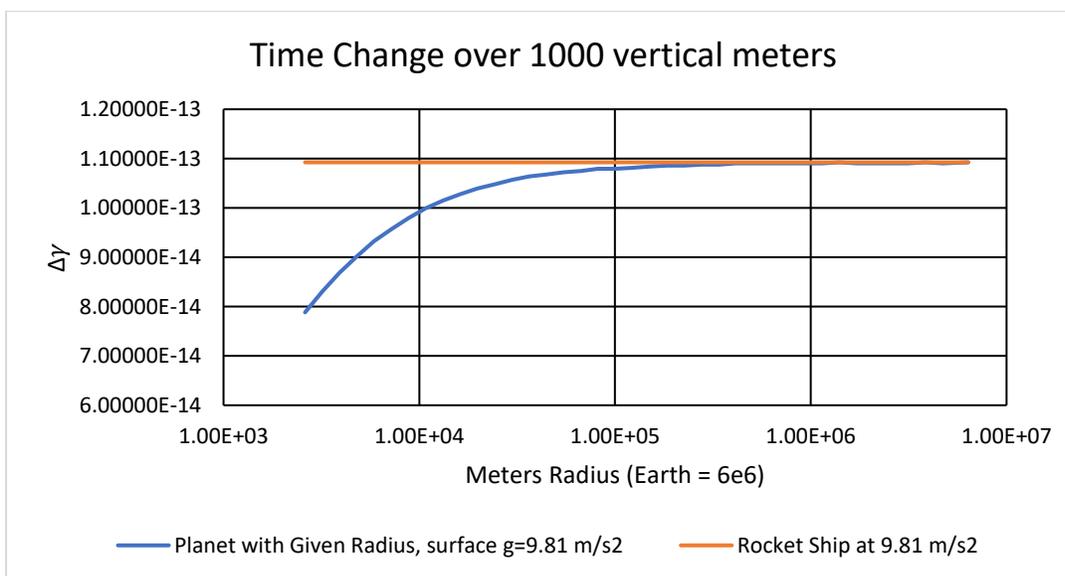

### The Bell Paradox Explained

This resolves the Bell Paradox directly. In a single spaceship, the time speed at the tip differs from that of the tail. A tail rider therefore would measure the acceleration as greater than that of a tip rider. That is, for the single spaceship to avoid compression or elongation, the tip and the tail must measure different accelerations. This is just as on Earth, where we measure different accelerations at different elevations.

The Bell Paradox scenario imposes the same acceleration at different elevations, and that will cause elongation. The lead spaceship will pull away. The reason is that "time runs faster up there" in an accelerating reference frame. It is possible to redefine the rules of motion for an accelerating reference frame; we do that and call it Rindler coordinates. Rindler coordinate systems typically involve hyperbolic math functions. They could tell us how to adjust acceleration of the following spacecraft to avoid breaking the string.

We can get to the same answer much easier up to significant speeds using Equation 4. Since the lead and tailing craft can agree on velocity but differ on time speed, their difference in terms of acceleration is the factor developed in Equation 4. That is, they can agree on a change in velocity when traveling from point A to B, but they disagree on the time by the calculated factor, so they disagree on acceleration (change in velocity/time) by the same factor. Specifically, for a 1000-meter separation and 9.81 m/s² acceleration of the lead spaceship, the tail craft should accelerate at 9.81 * (1+1.0925e-13) m/s², which is 9.81000000000107 m/s².

How can we test this? Use relativistic acceleration, start the two ships off, one with a 1000m head start, and do the math. Relativistic acceleration tells us that the acceleration we will measure, a, is less than that of the spaceship by a factor of $Υ^3$.

$$a = a'/Υ^3 \qquad [5]$$

Our calculated separation distances over time for a lead ship acceleration of 9.81 m/s² and a tailing ship acceleration of 9.81000000000107 m/s² does indeed show the ships slowing closing one on the other, but that is what we'd expect of a single ship's prow and tail. They get closer (by our measure) due to Lorentz contraction, our Equation 1. The graph of our calculated separation distance and the Lorentz contraction of a 1000m long spaceship (using 30,000 time steps out to 6 months and 0.55c) is:

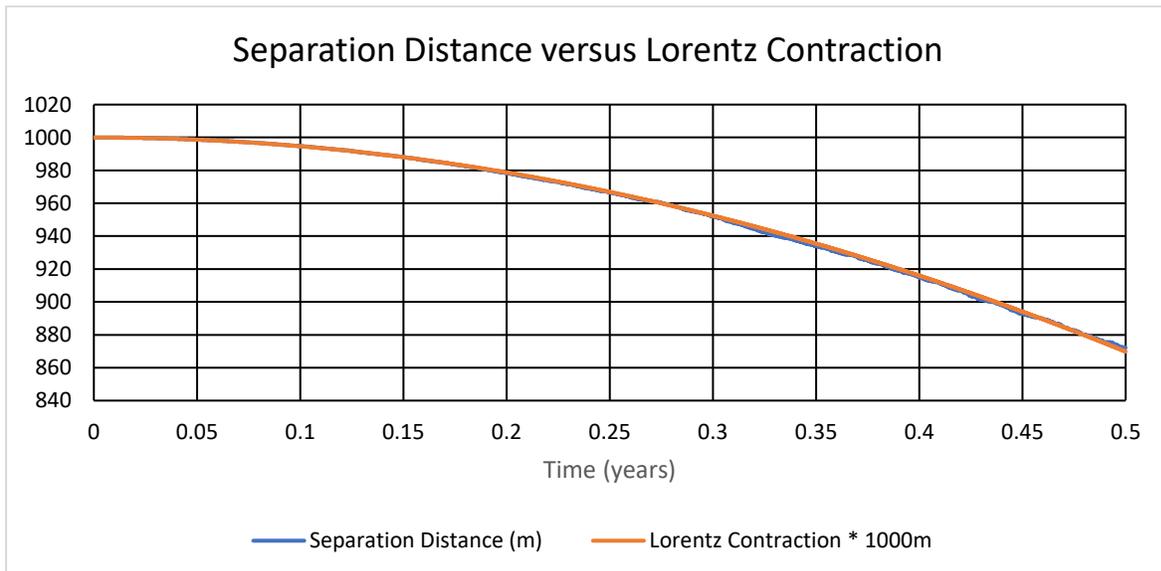

The agreement is not perfect, but we acknowledged an approximation in developing Equation 4, in that we used 1/2mv² for kinetic energy, and Equation 5 introduces approximations into a numeric analysis in which each time step is based on a constant Υ term. But our Equation 4 result is still good enough agreement to provide conceptual understanding of the Bell Paradox, and provides a parallel development to explain Lorentz contraction. The two string-connected ships with these different accelerations will not witness any decrease in separation, the string will not break or sag, yet we still viewers will calculate an approach just equal to the Lorentz contraction. If we try to impose equal accelerations on each ship, the pilots will see the other ship as having a different acceleration, the ships will pull apart, and the string will break not of an unusual relativistic electromagnetic stress, but of simple tension.

Note here that the idea of longer and shorter lengths is not just "faster is shorter." For our string-tied spaceships, we see their separation distance as shorter than the pilots calculate it, yet we see their traveled path through space as longer than the pilots calculate it.

An apt analogy is that of a space alien headed towards Earth from the North Star, viewing a six-foot-tall Santa Claus at the north pole. As the space alien accelerates to ever faster speed, if he wishes to keep Santa Claus's height at 6 feet, he must have an elf stretch Santa to a greater height as measured in the north pole reference frame. By the time $\Upsilon$=1.5 based on the alien velocity, Santa must be stretched to 9 feet to continue to appear 6 feet tall to the alien. The relative velocity does not cause a stress in Santa, but the stretching by the elf does. Similarly, the relative velocity between ourselves and the string-tied spaceships does not cause a stress, but forcing the separation distance to be constant in our eyes does cause a stress.

This concept can also be applied to circular paths. If we have a string of spaceships traveling in a circle at angles of π/6 radians (like the twelve digits on an analogue clock), as the velocity increases, each ship will calculate the distance to the leading and following ship as increasing. If tied with strings, the strings will break. This is the Ehrenfest paradox of a spinning disk; it is not possible to spin a disk up to relativistic velocities without the disk breaking apart. The reason, by our model, is that the imposed constant distance requires that the spaceship at 12:00 calculates a higher acceleration for the spaceship at 1:00, and a lower acceleration for the one at 11:00, hence velocity differences grow and the ships separate. We are attempting to stretch out the edge of the disk with the force being imposed by whatever is causing the rotation. The circular reference frame of the disk edge is being expanded by the calculation of the ship pilots. If we force them to remain at the coordinates as we view them, we will need to stretch them apart in some manner as the elf stretched Santa.

How can we test this? We can use the precession of satellites to check for consistency. Our model says that a satellite will calculate its 2πr path length as longer than we will (just as the string-tied ships calculated their separation distance as longer than we calculate it). The ratio of lengths is equal to the time factor, but because both gravity and velocity are involved, rather than the path length change being $\Upsilon$, the total path length difference is $\Upsilon\Upsilon_g$ with the terms as defined in Equations 1 and 3. Fortunately, orbital velocity is directly related to gravitational acceleration, so that we can express the path difference just in terms of $\Upsilon_g$. The math produces a time factor for objects in orbit of:

$$\Upsilon\Upsilon_g = (\Upsilon_g^2 + \Upsilon_g)/2 \quad \text{(for objects in orbit)} \quad [6]$$

For example, the International Space Station has a orbital radius of 6,783,000 meters ($\Upsilon_g$ = 1.000000000654 by Equation 3) and velocity of 7,666 m/s ($\Upsilon$ = 1.0000000003269 by Equation 1), providing a net time speed relative to still, deep space of 1.000000000981 by either side of Equation 6.

Since lengths change with time speed as in equation 1, the change in orbital path length is $2\pi r(\Upsilon\Upsilon_g-1)$. That is, the satellite doesn't complete a full revolution until it has traveled this extra distance, which causes an angle overshoot of $2\pi(\Upsilon\Upsilon_g-1)$ radians. Combining this with Equation 6 and performing some approximations gives a satellite precession per orbit of:

$$\varphi_s = 2\pi(\Upsilon\Upsilon_g - 1) = \pi(\Upsilon_g^2+\Upsilon_g - 2) \quad \text{or} \quad \varphi_s \approx \pi(\Upsilon_g^3 - 1) \quad \text{or} \quad \varphi_s \approx 1.5\pi r_s/r \quad [7]$$

The changed reference fame of the circular-path moving object is such that the object will travel an additional angle of $1.5\pi r_s/r$ radians each orbit. That is what is witnessed in experiments such as Gravity Probe-B, in which a very precise gyroscope is seen to tilt by 5.95e-5 radians per orbit (in the direction of an incomplete rotation). The math of Equation 7 is $1.5*\pi*0.0088681/7,021,000 = 5.95e-5$ radians. The approximations in developing Equation 7 are insignificant to be noticed out to several digits.

This measured change in tilt is proof of a measured change in orbit length. If no acceleration is involved, a second GP-B satellite can be tethered by string with the first and the string will not break no matter how long in orbit. If, however, through some machination, we accelerated the two GP-B satellites up to a much higher velocity while at the same radius, the overshoot in orbits will be proportionally larger, and the distance between satellites will grow, and the string will snap. Just like the edge of an Ehrenfest disk, the points separate, with motive power being whatever caused the satellite velocities to increase.

Simple relativity can similarly explain other relativity concepts, such as the Shapiro time delay, and the deflection of light by masses, in simple language . It can also resolve such problems as the twin astronaut paradox, or ladder in a barn paradox with similar simple concepts and mathematics. It is not a different model than conventional relativity, it is the same model viewed from a different approach, and it generates the same answers but with simpler formulas.

*Spreadsheets used to develop the figures are available from the author upon request, as is greater background in developing the math behind the equations.*

[1] Dewan, Edmond M.; Beran, Michael J. (March 20, 1959). "Note on stress effects due to relativistic contraction". *American Journal of Physics*. **27** (7): 517–518. Bibcode:1959AmJPh..27..517D. doi:10.1119/1.1996214.
[2] Evett, Arthur A.; Wangsness, Roald K. (1960). "Note on the Separation of Relativistically Moving Rockets". *American Journal of Physics*. **28** (6): 566. Bibcode:1960AmJPh..28..566E. doi:10.1119/1.1935893.
[3] *Dewan, Edmond M. (May 1963). "Stress Effects due to Lorentz Contraction". American Journal of Physics. **31** (5): 383–386. Bibcode:1963AmJPh..31..383D. doi:10.1119/1.1969514.* (Note that this reference also contains the first presentation of the ladder paradox.)
[4] *Romain, Jacques E. (1963). "A Geometrical Approach to Relativistic Paradoxes". American Journal of Physics. **31** (8): 576–585. Bibcode:1963AmJPh..31..576R. doi:10.1119/1.1969686.*
[5] *Evett, Arthur A. (1972). "A Relativistic Rocket Discussion Problem". American Journal of Physics. **40** (8): 1170–1171. Bibcode:1972AmJPh..40.1170E. doi:10.1119/1.1986781.*
[6] *Gershtein, S. S.; Logunov, A. A. (1998). "J. S. Bell's problem". Physics of Particles and Nuclei. **29** (5): 463–468. Bibcode:1998PPN....29..463G. doi:10.1134/1.953086.*
[7] *Tartaglia, A.; Ruggiero, M. L. (2003). "Lorentz contraction and accelerated systems". European Journal of Physics. **24** (2): 215–220. arXiv:gr-qc/0301050. Bibcode:2003EJPh...24..215T. doi:10.1088/0143-0807/24/2/361.*
[8] *Cornwell, D. T. (2005). "Forces due to contraction on a cord spanning between two spaceships". EPL. **71** (5): 699–704. Bibcode:2005EL......71..699C. doi:10.1209/epl/i2005-10143-x.*
[9] *Flores, Francisco J. (2005). "Bell's spaceships: a useful relativistic paradox". Physics Education. **40** (6): 500–503. Bibcode:2005PhyEd..40..500F. doi:10.1088/0031-9120/40/6/F03.*

[10] *Semay, Claude (2006). "Observer with a constant proper acceleration". European Journal of Physics.* **27** *(5): 1157–1167.* arXiv:physics/0601179. Bibcode:2006EJPh...27.1157S. doi:10.1088/0143-0807/27/5/015.

[11] *Styer, Daniel F. (2007). "How do two moving clocks fall out of sync? A tale of trucks, threads, and twins". American Journal of Physics.* **75** *(9): 805–814.* Bibcode:2007AmJPh..75..805S. doi:10.1119/1.2733691.

[12] *Jürgen Freund (2008). "The Rocket-Rope Paradox (Bell's Paradox)". Special Relativity for Beginners: A Textbook for Undergraduates. World Scientific. pp. 109–116.* ISBN 978-9812771599.

[13] *Redžić, Dragan V. (2008). "Note on Dewan Beran Bell's spaceship problem". European Journal of Physics.* **29** *(3): N11–N19.* Bibcode:2008EJPh...29...11R. doi:10.1088/0143-0807/29/3/N02.

[14] *Peregoudov, D. V. (2009). "Comment on 'Note on Dewan-Beran-Bell's spaceship problem'". European Journal of Physics.* **30** *(1): L3–L5.* Bibcode:2009EJPh...30L...3P. doi:10.1088/0143-0807/30/1/L02.

[15] *Redžić, Dragan V. (2009). "Reply to 'Comment on "Note on Dewan-Beran-Bell's spaceship problem"'". European Journal of Physics.* **30** *(1): L7–L9.* Bibcode:2009EJPh...30L...7R. doi:10.1088/0143-0807/30/1/L03.

[16] *Gu, Ying-Qiu (2009). "Some Paradoxes in Special Relativity and the Resolutions". Advances in Applied Clifford Algebras.* **21** *(1): 103–119.* arXiv:0902.2032. doi:10.1007/s00006-010-0244-6.

[17] Vesselin Petkov (2009): Accelerating spaceships paradox and physical meaning of length contraction, arXiv:0903.5128, published in: *Veselin Petkov (2009). Relativity and the Nature of Spacetime. Springer.* ISBN 978-3642019623.

[18] Franklin, Jerrold (2010). "Lorentz contraction, Bell's spaceships, and rigid body motion in special relativity". *European Journal of Physics*. **31** (2): 291–298. arXiv:0906.1919. Bibcode:2010EJPh...31..291F. doi:10.1088/0143-0807/31/2/006.

[19] *Miller, D. J. (2010). "A constructive approach to the special theory of relativity". American Journal of Physics.* **78** *(6): 633–638.* arXiv:0907.0902. Bibcode:2010AmJPh..78..633M. doi:10.1119/1.3298908.

[20] *Fernflores, Francisco (2011).* "Bell's Spaceships Problem and the Foundations of Special Relativity". *International Studies in the Philosophy of Science.* **25** *(4): 351–370.* doi:10.1080/02698595.2011.623364.

[21] *Kassner, Klaus (2012). "Spatial geometry of the rotating disk and its non-rotating counterpart". American Journal of Physics.* **80** *(9): 772–781.* arXiv:1109.2488. Bibcode:2012AmJPh..80..772K. doi:10.1119/1.4730925.

[22] *Natario, J. (2014). "Relativistic elasticity of rigid rods and strings". General Relativity and Gravitation.* **46** *(11): 1816.* arXiv:1406.0634. doi:10.1007/s10714-014-1816-x.

[23] *Lewis, G. F., Barnes, L. A., & Sticka, M. J. (2018). "Bell's Spaceships: The Views from Bow and Stern". Publications of the Astronomical Society of Australia.* **35***: e001.* arXiv:1712.05276. Bibcode:2018PASA...35....1L. doi:10.1017/pasa.2017.70.

[24] *Bokor, N. (2018). "Playing Tag Relativistically". European Journal of Physics.* **39** *(5): 055601.* Bibcode:2018EJPh...39e5601B. doi:10.1088/1361-6404/aac80c.

[25] Forshaw, Jeffrey; Smith, Gavin (2014). *Dynamics and Relativity*. John Wiley & Sons. ISBN 978-1-118-93329-9.